\documentclass[10pt]{article}
\pdfoutput=1
\usepackage{graphicx}
\usepackage[square,numbers,sort&compress]{natbib}
\usepackage{subfig}

\newcommand{\ie}{\emph{i.e.},~}

\newcommand{\cf}{\emph{cf.~}}

\title{Mesoscale modelling of polyelectrolyte electrophoresis}

\author{Kai Grass$^{a}$ and Christian Holm$^{a,b,c}$\\[3mm]
$^a$ Frankfurt Institute for Advanced Studies, Goethe University,\\[1mm]
Ruth-Moufang-Str. 1, 60438 Frankfurt/Main, Germany.\\[1mm]
E-mail: grass@fias.uni-frankfurt.de\\[1mm]
$^b$ Max-Planck-Institut f\"ur Polymerforschung, \\[1mm]
Ackermannweg 10, 55128 Mainz, Germany. \\[1mm]
$^c$  Institute for Computational Physics, University of Stuttgart, \\[1mm]
Pfaffenwaldring 27, 70569 Stuttgart,\\[1mm]
Germany.\\[1mm]
E-mail: holm@icp.uni-stuttgart.de}

\begin{document}

\maketitle

\renewcommand{\thefootnote}{\fnsymbol{footnote}}

\noindent The electrophoretic behaviour of flexible polyelectrolyte chains
ranging from single mono\-mers up to long fragments of hundred repeat units is
studied by a mesoscopic simulation approach. Abstracting from the atomistic
details of the polyelectrolyte and the fluid, a coarse-grained molecular dynamics
model connected to a mesoscopic fluid described by the Lattice Boltzmann approach
is used to investigate free-solution electrophoresis. Our study demonstrates the
importance of hydrodynamic interactions for the electrophoretic motion of
polyelectrolytes and quantifies the influence of surrounding ions. The
length-dependence of the electrophoretic mobility can be understood by evaluating
the scaling behavior of the effective charge and the effective friction. The
perfect agreement of our results with experimental measurements shows that all
chemical details and fluid structure can be safely neglected, and a suitable
coarse-grained approach can yield an accurate description of the physics of the
problem, provided that electrostatic and hydrodynamic interactions between all
entities in the system, \ie the polyelectrolyte, dissociated counterions,
additional salt and the solvent, are properly accounted for. Our model is able to
bridge the single molecule regime of a few nm up to macromolecules with contour
lengths of more than 100 nm, a length scale that is currently not accessible to
atomistic simulations.

\section{Introduction}\label{sec:introduction}

Nowadays, electrophoresis methods are widely used to separate
biomolecules~\cite{righetti96a,dolnik06a} such as peptides, proteins, DNA, as
well as synthetic polymers~\cite{cottet05a,cottet07a}. In order to be able to
improve the processes involved in current electrophoretic separation methods it
is a prerequisite to gain a thorough understanding of the behaviour of
polyelectrolytes (PEs) in an externally applied electric field. Several
theories~\cite{barrat96a,muthukumar96a,volkel95b,mohanty99a} have been developed
to describe PE electrophoresis and successfully described qualitatively the
experimentally observed behaviour of various PEs under bulk conditions. However,
the mobility of small oligomeric PEs showed under low salt conditions a
non-monotonic behaviour that current theories had not been able to explain.

% \begin{figure}[htp]
% \begin{center}
%   \includegraphics[width=\columnwidth]{mobility-withinset}
%   \label{fig:pss-mobility}
%   \caption{The normalized electrophoretic mobility
%     $\mu / \mu_{\mathrm{FD}}$ as a function of the number of repeat
%     units $N$ for simulation data including hydrodynamic interactions (HI),
%     and experimental data coming from capillary electrophoresis (CE) and from
%     electrophoretic NMR.  The inset compares to simulation data obtained with a
%     model neglecting hydrodynamic interactions.}
% \end{center}
% \end{figure}

\begin{figure}[htp]
\begin{center}
  \includegraphics[width=\columnwidth]{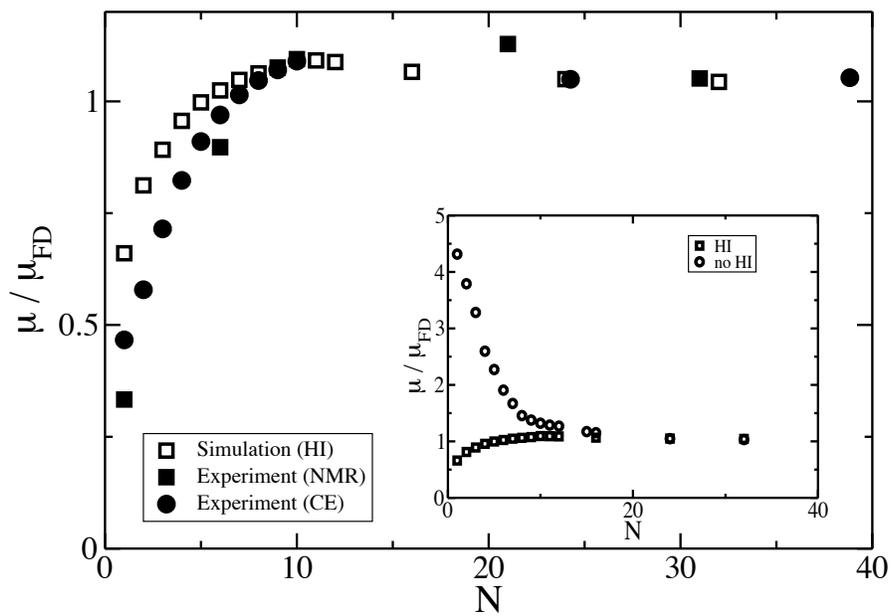}
  \label{fig:pss-mobility}
  \caption{The normalized electrophoretic mobility
    $\mu / \mu_{\mathrm{FD}}$ as a function of the number of repeat
    units $N$ for simulation data including hydrodynamic interactions (HI),
    and experimental data coming from capillary electrophoresis (CE) and from
    electrophoretic NMR.  The inset compares to simulation data obtained with a
    model neglecting hydrodynamic interactions.}
\end{center}
\end{figure}

In a recent publication~\cite{grass08a}, we employed a mesoscopic coarse-grained
model using molecular dynamics simulations in connection with a Lattice-Boltzmann
(LB) algorithm to extend the theoretical understanding on a more detailed level,
and in particular, we intended to investigate the role of hydrodynamic
interactions in these systems. Our results were able to match the free-solution
electrophoretic mobility $\mu$ of short polyelectrolyte chains, here polystyrene
sulfonate (PSS), as a function of the number of repeat units $N$ with
quantitative agreement to experiments as shown in Figure~\ref{fig:pss-mobility}.
Since the three data sets have different solvent viscosities the mobility is
normalized by the corresponding constant mobility for long chains, the so-called
free-draining mobility, $\mu_\mathrm{FD}$. The electrophoretic mobility increases
for short oligomers, reaches a maximum for intermediate degrees of
polymerization, and slowly decreases towards a plateau value for long chains. To
understand this observation, the hydrodynamic interactions were investigated in
detail and we found that they are actually the major driving force for the length
dependent mobility for short and intermediate chain lengths. The constant
mobility for long chains can be attributed to an effective screening of
hydrodynamic interactions, which leads to the so-called free-draining behavior.
The inset in Figure~\ref{fig:pss-mobility} shows a comparison to a coarse-grained
simulation that neglects hydrodynamic interactions. This leads to a qualitatively
completely different behavior, showing a monotonically decreasing mobility.
Agreement to the experimentally observed behaviour is only achieved as long as
hydrodynamic interactions are included correctly as has been shown in detail in
our previous investigations~\cite{grass08c,grass08b}.

In this article, we will extend our work by studying the electrophoresis of
generic flexible polyelectrolyte chains ranging from single mono\-mers to
long fragments of hundred repeat units. Abstracting from the atomistic details of
the polyelectrolyte and the fluid, a coarse-grained molecular dynamics model
connected to a mesoscopic fluid described by the Lattice Boltzmann approach is
used to investigate the free-solution behavior under varying salt concentration.

In the next section we will introduce the employed simulation model. In
Section~\ref{sec:results}, the main results of this study are presented and
discussed. We conclude with final remarks in Section~\ref{sec:conclusions}.

\section{Model}\label{sec:model}

We employ molecular dynamics (MD) simulations using the ESPResSo
package~\cite{limbach06a} to study the behaviour of linear polyelectrolytes
(PE) of different lengths. The PEs are modelled by a totally flexible
bead-spring model. The monomers are connected to each other by finitely
extensible nonlinear elastic (FENE) bonds~\cite{soddeman01a}
\[
U_\mathrm{FENE}(r) = \frac{1}{2} k R^2 \ln \left( 1 - \left( \frac{r}{R} \right)^2
\right),
\]
with stiffness $k = 30 \epsilon_0$, and maximum extension $R = 1.5
\sigma_0$, where $r$ is the distance between the interacting monomers.
Additionally, a truncated Lennard-Jones or WCA potential~\cite{weeks71a}
\[
U_\mathrm{LJ}(r<r_\mathrm{c}) = \epsilon_0 \left( \left(
\frac{\sigma_0}{r}\right)^{12} - \left(
\frac{\sigma_0}{r}\right)^6 + \frac{1}{4} \right),
\]
is used for excluded volume interactions between all monomers. A cutoff value
of $r_c = \sqrt[6]{2}\sigma_0$ ensures a purely repulsive potential. All
dissociated counterions and additional salt ions are modelled by appropriately
charged spheres using the same WCA potential.

Here, $\epsilon_0$ and $\sigma_0$ define the energy and length scale of the
simulations. We use $\epsilon_0 = k_\mathrm{B} T$, i.e.~the energy of the system is
expressed in terms of the thermal energy. The length scale $\sigma_0$ defines the
size of the monomers and the dimension of the system. For this study, $\sigma_0$ is
chosen to be $4~\mathrm{\AA}$. Different polyelectrolytes can be mapped by changing
$\sigma_0$. Unless mentioned otherwise, all observables are expressed in reduced
simulation units, and we will not use $\sigma_0$ and $\epsilon_0$ explicitly from
now on.

The chain length is varied from $N=1$ to $N=128$ and all chain monomers carry a
negative electric charge $q = -1 e_0$, where $e_0$ is the elementary charge. For
charge neutrality, $N$ monovalent counterions of charge $+1 e_0$ are added.
Additional monovalent salt is added to the simulation, corresponding to
concentrations between $c_\mathrm{s} = 0 \mathrm{~mM}$ and $c_\mathrm{s} = 160
\mathrm{~mM}$. The later concentration being equivalent to a particle density of
the salt ions of $\rho_\mathrm{s} = 0.01$.

A homogeneous electric field with reduced field strength $E=0.1$ is applied in
x-direction creating a force $F_\mathrm{E}=q E$ on all charged particles, and
thus inducing an electrophoretic mobility. It has been carefully checked that the
field strength is within the linear response regime, \ie it does not influence
the chain conformation or the distribution of the surrounding
ions~\cite{grass08b}.

Full electrostatic interactions are calculated with the P3M algorithm using the
implementation of Reference~\citealp{deserno98a}. The Bjerrum length
\[
  l_\mathrm{B} = e_0^2 / \left( 4 \pi \epsilon_0 \epsilon_\mathrm{r} k_\mathrm{B}
  T \right)= 1.8
\]
in simulation units corresponds to 7.1 \AA, the Bjerrum length in water at
room temperature. This means that the effect of the surrounding water is modelled
implicitly by simply using the dielectric properties of water, having a relative
dielectric constant of $\epsilon_r \approx 80$.

The simulations are carried out under periodic boundary conditions in a cubic
simulation box. The size $L$ of the box is varied to realize a constant monomer
concentration of $c_\mathrm{PE} = 16 \mathrm{~mM}$ independent of chain length.
This is equivalent to a monomer density $\rho_\mathrm{PE} = 0.001$.

We include hydrodynamic interactions by using a Lattice Boltzmann 
algorithm~\cite{mcnamara88a} that is interacting with the beads in the MD
simulations via a frictionally coupling introduced by Ahlrichs et
al.~\cite{ahlrichs99a}. The mesoscopic LB fluid is described by a velocity
field generated by discrete momentum distributions on a spatial grid rather
than explicit fluid particles. We use an implementation of the D3Q19 model
with a kinematic viscosity $\nu = 1.0$, and a fluid density $\rho = 1.0$. The
resulting fluid has a dynamic viscosity $\eta = \rho \nu = 1.0$. The
simulation box is discretised by a grid with spacing $a = 1.0$. As usual in a
standard Langevin approach, the particle-fluid interaction is realised by a
dissipative force. This force depends on the difference between the particle
velocity $\mathbf{v}$ and the fluid velocity at the particle position
$\mathbf{u}$:
\[
  \mathbf{F_\mathrm{R}} = -\Gamma_\mathrm{bare} (\mathbf{v}-\mathbf{u}).
\]
Here, the coupling constant takes on the value $\Gamma_\mathrm{bare} = 20.0$.
Additional random fluctuations introduced to the particles and fluid act as a
thermostat. The interaction between particles and fluid conserve total
momentum, and this algorithm has been shown to yield correct long-range
hydrodynamic interaction between individual particles~\cite{ahlrichs99a}.

Additionally, a second type of MD simulation is used which is based on the
Langevin equations of motions with a velocity dependent dissipative and a random
term in addition to the inter particle forces. Together, both additional terms
implicitly model the effects of a solvent surrounding the particles: the
dissipative force,
\[
  \mathbf{F_\mathrm{D}} = -\Gamma_0 \mathbf{v},
\]
with $\Gamma_0 = 1.0$ provides local friction and the non-correlated zero-mean
Gaussian random forces,
\[
  \mathbf{F_\mathrm{R}} = \mathbf{\xi}(t),
\]
mimic thermal kicks (Brownian motion). In order to fulfill the
fluctuation-dissipation theorem, dissipative and random force have to be
coupled together: $\langle \xi_i(t) \cdot \xi_j(t') \rangle = 6 \Gamma_0
k_\mathrm{B} T \delta_{ij} \delta(t-t')$. This approach only offers local
particle-fluid interactions, and therefore destroys long-range hydrodynamic
interactions. Nevertheless one can use it to compute the effective charge as
has been presented in~\cite{grass08c,grass08b}. This effective charge is used
to obtain the effect friction in the presence of hydrodynamic interactions and
illustrates their importance for the electrophoretic mobility.

All simulations are carried out with a MD time step $\tau_\mathrm{MD} = 0.01$ and
LB time step $\tau_\mathrm{LB} = 0.05$. After an equilibration time of $10^6$
steps, $10^7$ steps are used for generating the data. The time-series of four
independent simulations are analyzed using auto-correlation functions to estimate
the statistical errors as detailed in Reference~\citealp{wolff04a}. Error bars
of the order of the symbol size or smaller are omitted in the figures.

\section{Results and discussion}\label{sec:results}

\subsection{Electrophoretic mobility}

We determine the electrophoretic mobility $\mu$ as the ratio between the
measured center of mass velocity $v_\mathrm{PE}$ and the magnitude of the
electric field $E$:
\[
  \mu = \frac{v}{E}.
\]
For comparison, the results are normalized by the monomer mobility $\mu_1$.

% \begin{figure}[htp]
% \begin{center}
%   \includegraphics[width=\columnwidth]{mobility-hd}
%   \caption{The normalized electrophoretic mobility $\mu/\mu_1$ of polyelectrolyte chains of length
%   $N$ for three different salt concentrations using the LB algorithm. The added salt not only influences
%   the absolute mobility, but likewise changes the characteristic shape of the
%   mobility with respect to chain length $N$.}
%   \label{fig:mobility-hd}
% \end{center}
% \end{figure}

Figure~\ref{fig:mobility-hd} displays the characteristic behaviour of flexible
polyelectrolytes for vanishing salt concentration $c_\mathrm{s}=0 \mathrm{~mM}$:
initially, the electrophoretic mobility increases with $N$ to reach a maximum at
intermediate chain lengths and then slowly decays towards a constant value for
long chains. This constant value, often called the free-draining limit
$\mu_\mathrm{FD}$, can be explained by the length independence of the ratio
between effective charge and effective friction for long chains as we will show
in this article.

In the presence of added salt, the long chain mobility is reduced, which is
consistent with the experimentally observed~\cite{hoagland99a} behavior.
Furthermore, the shape of the curve is influenced, and the maximum at
intermediate chains is suppressed for increased salt concentration. At
$c_\mathrm{s}=160 \mathrm{~mM}$ the maximum disappears and the measured mobility
becomes length independent within the resolution of the simulation. A further
increase of the added salt concentration leads to a further reduction of the
limiting mobility $\mu_\mathrm{FD}$ , not shown here, while the monomer mobility
$\mu_1$ remains almost unchanged. This leads eventually to an inverted
length-dependence with a monotonic decrease of the mobility towards the limiting
value.

In the simple local force picture, the constant center of mass velocity $v_\mathrm{PE}$
that determines the electrophoretic mobility is a direct result of the cancellation of two acting
forces: the electric driving force $F_\mathrm{E}= Q_\mathrm{eff} E$ is canceled
by the solvent friction or drag force $F_\mathrm{D} = \Gamma_\mathrm{eff}
v_\mathrm{PE}$. Here, $Q_\mathrm{eff}$ is the effective charge of the
polyelectrolyte, which can be thought of as the bare charge of the polyelectrolyte reduced by
oppositely charged ions in solution that associate to the polyelectrolyte chain.
The association of counterions to a PE chain is known as counterion
condensation~\cite{manning69a,oosawa71a}. The compound formed by the polyelectrolyte
and the associated ions is moved through the solvent under the influence of the
external field and experiences a Stokesian drag force with an effective friction
coefficient $\Gamma_\mathrm{eff}$ that is a priori unknown. In the steady state
both forces balance and the mobility is given by
\[
  \mu = \frac{v}{E} = \frac{Q_\mathrm{eff}}{\Gamma_\mathrm{eff}}.
\]

% \begin{figure}[htp]
% \begin{center}
%   \includegraphics[width=\columnwidth]{mobility-nohd}
%   \caption{The normalized electrophoretic mobility $\mu/\mu_1$ for different chain
%   length $N$ at varying salt concentrations $c_\mathrm{s}$ without hydrodynamic
%   interactions differs significantly from the behaviour observed in
%   Figure~\ref{fig:mobility-hd}. The mobility shows a salt-dependent monotonic
%   decrease for short chains and a salt-independent constant value for long
%   chains.}
%   \label{fig:mobility-nohd}
% \end{center}
% \end{figure}

Next, let us compare the results of Figure~\ref{fig:mobility-hd} to the case when
long-range hydrodynamic interactions between the particles are neglected in
simulations, \ie by using a standard Langevin thermostat. The results are shown
in Figure~\ref{fig:mobility-nohd}. One immediately notices that the observed
electrophoretic mobility differs significantly from the behaviour observed in
Figure~\ref{fig:mobility-hd}. Independent of the salt concentration, the mobility
decreases monotonically with chain length and slowly approaches a constant value
for long chains which is independent of the salt concentration. This
difference to the experimental observations and to the LB simulation including
hydrodynamics will be analyzed in detail in the following sections.

\subsection{Effective charge}

To analyze the observed influence of the added salt on the polyelectrolyte
mobility, we will determine the effective charge, and can then calculate
$\Gamma_\mathrm{eff}=Q_\mathrm{eff}/\mu$ to obtain an estimate for the effective
friction of the polyelectrolyte-ion compound. A word of care has to be taken
here, since the value of the effective charge depends on definition.
Qualitatively one can differentiate between a static definition and a dynamic
definition~\cite{wette02a}. In our case it is obviously a dynamic definition.
In~\cite{grass08c,grass08b}, we introduced several static and dynamic estimators
for the effective charge and showed their equivalence at vanishing salt
concentration. Here, three of them will be reviewed and applied to the case of
added salt.

The local force picture described above can be used to estimate the effective
charge of the polyelectrolyte based on the measurement of the electrophoretic
mobility in the absence of hydrodynamic interactions. Let $N_\mathrm{CI}$ be the
number of associated counterions reducing the bare charge of the polyelectrolyte
which is equal to $N$. The effective charge is then given by
\[
  Q_\mathrm{eff} = N - N_\mathrm{CI}.
\]
Without long-rang hydrodynamic interactions the interaction of each particle
with the solvent is purely local and directly given by the friction constant
$\Gamma_0$ of the Langevin algorithm. The total effective friction of the
polyelectrolyte and the ions is then:
\[
 \Gamma_\mathrm{eff} = \Gamma_0 \left( N + N_\mathrm{CI} \right).
\]
This results in an expression for the electrophoretic mobility
\[
  \mu = \frac{N-N_\mathrm{CI}}{\Gamma_0 \left( N + N_\mathrm{CI}\right)}
\]
from which an expression for $N_\mathrm{CI}$ is obtained. Therefore we 
can express the effective charge purely as a function of the mobility 
measurements shown in Figure~\ref{fig:mobility-nohd} and on our input 
value for $\Gamma_0$, independent of the knowledge of the value of $N_cl$ by the following
expression:
\[
  Q_\mathrm{eff}^{(1)} = N \left( 1 - \frac{1-\mu \Gamma_0}{1+\mu \Gamma_0}\right).
\]

% \begin{figure}[htp]
% \begin{center}
%   \includegraphics[width=\columnwidth]{velcor}
%   \caption{The average ion velocity in the direction of the electric field $v_\mathrm{CI}$ (here for
%   a chain with $N=64$ monomers, salt concentration $c_\mathrm{s}=16\mathrm{~mM}$, and hydrodynamics included)
%   depends on the distance $d$ to the center of mass of the polyelectrolyte. Ions
%   close to the center co-move with the chain's velocity
%   (dashed line), whereas ions far away from the center move with the single particle
%   velocity $v_1 = \mu_1 E$ into the opposite direction. The distance $d_0$ at which
%   $v_\mathrm{CI}\left(d_0\right)=0$ is used to separate co-moving, associated
%   ions from non-associated ones.
%   The solid line shows the integrated fraction of charges $I$ that is found up to
%   the distance $d$ of the center of mass.}
%   \label{fig:velcor}
% \end{center}
% \end{figure}

An alternative way of characterizing the associated ions is presented in
Ref.~\citealp{lobaskin04b} who suggested to determine the ion velocity with
respect to the distance to the center of mass of the polyelectrolyte. For this
method we use the LB algorithm to include hydrodynamical interactions, and the
result for a chain of $N=64$ at $c_s = 16 \mathrm{~mM}$ can be inspected in
Figure~\ref{fig:velcor}. The average ion velocity in the direction of the
electric field $v_\mathrm{CI}$ is a function of the distance $d$ to the center
of mass of the polyelectrolyte chain and in general depends on the chain length
$N$ and the salt concentration $c_\mathrm{s}$. As shown, ions close to the
center move with the chain at negative speed, whereas ions far away from
the center move with the single particle velocity $v_1 = \mu_1 E$ into the
opposite direction. The association of ions to the chain is strong enough to
move them against the electric field. For every chain length and every salt
concentration, the distance $d_0$ at which $v_\mathrm{CI}\left(d_0\right)=0$ is
used to separate co-moving, associated ions from non-associated ones.

We use this distance $d_0$ to define the effective charge by summing up the total
charge in the system found within this distance to the center of mass of the polyelectrolyte:
\[
  Q_\mathrm{eff}^{(2)} = N \left(1-I\left(d_0\right)\right),
\]
where $I\left(d_0\right)$ is the integrated fraction of neutralizing charges found by
adding the number of counterions and positively charged salt ions reduced by the number
of negatively or like-charged salt ions. Far away from the center of mass of the chain,
the total bare charge of the polyelectrolyte is neutralized and $I=1$.

% \begin{figure}[htp]
% \begin{center}
%   \subfloat[]{
%     \includegraphics[width=0.45\columnwidth]{qeff}
%     \label{fig:qeffa}
%   }
%   \subfloat[]{
%     \includegraphics[width=0.45\columnwidth]{qeffN}
%     \label{fig:qeffb}
%   }
%   \caption{(a) The effective charge $Q_\mathrm{eff}$ as a function of chain
%    length $N$ (symbols for $Q_\mathrm{eff}^{(1)}$, lines for $Q_\mathrm{eff}^{(2)}$).
%    Both charge estimators show good agreement. Initially, $Q_\mathrm{eff}$ is
%    close to the bare charge $N$ (dotted line), but as ion condensation sets in,
%    the effective charge is reduced. Longer
%    chains show a linear increase of their charge close to the Manning prediction
%    $\left( {1}/{\xi}\right) N$ (dashed line).
%    (b) The effective charge per monomer $Q_\mathrm{eff/N}$ is influenced by
%    the salt concentration for short and intermediate chains. The higher the
%    concentration of the added salt, the faster the electric charge of the
%    polyelectrolyte is reduced by condensed ions. For long chains, the charge
%    per monomer is again independent of the salt concentration and comparable to
%    the Manning prediction $1/\xi$ (dashed line).}
%   \label{fig:qeff}
% \end{center}
% \end{figure}

The effective charge $Q_\mathrm{eff}$ as obtained from both estimators is
presented in Figure~\ref{fig:qeffa}. Initially, $Q_\mathrm{eff}$ is close to the
bare charge $N$, but as ion condensation sets in, the effective charge is
reduced. Longer chains show a linear increase of their charge close to the
Manning prediction for counterion condensation in the salt free case
$Q_\mathrm{eff} = \left({1/\xi}\right) N$, where Manning parameter $\xi
= {l_\mathrm{B}}/b$ is the ratio between the Bjerrum length and the charge spacing
along the polyelectrolyte backbone. For the model used here $b=0.9$ and therefore
$\xi = 2.0$. We note that there is no apparent dependence of the effective
charge for long polyelectrolyte chains on the salt concentrations
when measured by the dynamic effective charge estimators presented here.

Figure~\ref{fig:qeffb} plots the effective charge per monomer,
$Q_\mathrm{eff}/N$. Here, the influence of the salt concentration for short and
intermediate chain length can be seen. The higher the concentration of the added
salt, the faster the electric charge of the polyelectrolyte is reduced by
condensed counter ions. For long chains, the charge per monomer is again
independent of the salt concentration and comparable to the Manning prediction
$1/\xi$. The difference for short and intermediate chains at different salt concentrations
can be attributed to stronger association of counterions with increasing salt
concentrations. For short chains, smaller than the Debye length, effects due to
the finite size play a leading role in the ability to condense
counterions~\cite{limbach01a,antypov06a,antypov07a}.

Additionally, Figure~\ref{fig:qeff} shows the equivalence of the two
dynamic estimators $Q_\mathrm{eff}^{(1)}$ and $Q_\mathrm{eff}^{(2)}$
independently of the presence or absence of hydrodynamic interactions also in
the presence of additional salt. This new observation supports the
applicability and importance of these charge estimators for the study of
polyelectrolytes during electrophoresis. 

The charge estimators $Q_\mathrm{eff}^{(1)}$ and $Q_\mathrm{eff}^{(2)}$ measure
the effective charge of the moving polyelectrolyte and its surrounding
counterions. Therefore, they measure the effective dynamic charge of the
polyelectrolyte. Similarly, it is possible to define a static estimate of the
effective charge using the following simple method
\begin{displaymath}
	Q_\mathrm{eff}^{(3)} = N_\mathrm{PE} - N_\mathrm{CI}(d<d_0),
\end{displaymath}
where $N_\mathrm{CI}(d<d_0)$ is the average number of counterions that can be
found within a distance $d$ to the closest monomer. Here, the threshold $d_0$ 
chosen to be $d_0 = 2 \sigma_0$.

The second static charge estimator used in Ref.~\citealp{grass08b} based on the
inflection criterion to estimate the threshold of counterion
condensation~\cite{belloni84a,belloni98a,deserno00a} breaks down in the
presence of high salt concentrations and therefore can not be applied here.

% \begin{figure}[htp]
% \begin{center}
%   \includegraphics[width=\columnwidth]{qeffstatic}
%   \caption{The effective charge as measured by the static estimator
%   $Q_\mathrm{eff}^{(3)}$ shows a strong dependence on the the salt
%   concentration. At $c_\mathrm{S} = 0 \mathrm{~mM}$ the static estimate
%   agrees with the dynamic estimates (solid line). The higher the salt
%   concentration, the lower is the static charge estimate. For comparison the
%   bare charge $N$ (dotted line) and the Manning prediction (dashed-line) are plotted.}
%   \label{fig:qeffstatic}
% \end{center}
% \end{figure}

In Figure~\ref{fig:qeffstatic}, we compare the static charge estimate
$Q_\mathrm{eff}^{(3)}$ for varying salt concentrations to the dynamic charge
estimate obtained by $Q_\mathrm{eff}^{(1)}$. Unlike the dynamic estimate the
static charge estimate shows a strong dependence on the salt concentration.
While both estimators agree for vanishing salt concentration as previously
shown in Ref.~\citealp{grass08b}, the static charge estimate shows a decreases
with higher salt concentrations, hence an increase in counterion
condensation, as could be expected from a mean-field
comparison~\cite{deserno00a}, and eventually falls below the Manning prediction.

The higher salt concentrations increase the number of counterions in the close
vicinity of the chain as measured by $Q_\mathrm{eff}^{(3)}$. At the same time,
the electrostatic interactions in the system are reduced due to electrostatic
screening, which also reduces the strength of the coupling between the
polyelectrolyte and the counterions. The independence of the dynamic effective
charge on the salt concentration for long chains as shown in
Figure~\ref{fig:qeff} has to be understood as the cancellation of both effects:
with increasing salt concentration more counterions in close vicinity to the
polyelectrolyte are influenced by the chain but the strength of the
interactions is reduced in such a way that the combined action remains
unchanged and yields a concentration independent dynamic effective charge.

In the following, we will use the dynamic effective charge to calculate
the effective friction of the polyelectrolyte-ion compound.

\subsection{Effective friction}\label{sec:results-friction}

When long-range hydrodynamic interactions are present, the effective friction
of the polyelectrolyte and the associated counterions cannot be given in
a simple analytic form. We therefore obtain it from the measurements of the
mobility and the effective charge presented above:
\[
  \Gamma_\mathrm{eff} = \frac{Q_\mathrm{eff}}{\mu}.
\]

% \begin{figure}[htp]
% \begin{center}
%   \subfloat[]{
%     \includegraphics[width=0.45\columnwidth]{gammaeff}
%     \label{fig:gammaeffa}
%   }
%   \subfloat[]{
%     \includegraphics[width=0.45\columnwidth]{gammaeffN}
%     \label{fig:gammaeffb}
%   }
%   \caption{(a) The normalized effective friction $\Gamma_\mathrm{eff} \mu_1$ as a function of
%   chain length $N$ for different salt concentrations $c_\mathrm{s}$ using the LB algorithm. Initially,
%   the friction increases as given by the hydrodynamic size of the polyelectrolyte
%   $\Gamma = \sim N^{0.59}$ (dotted line). With the onset of counterion
%   condensation the friction exceeds the value of the bare polyelectrolyte
%   and for long chains becomes linear in
%   $N$ (dashed line). The absolute friction value is increased with the addition of external salt.
%   (b) The normalized effective friction per monomer $\Gamma_\mathrm{eff}\mu_1/N$
%   shows an initial decrease with chain length that is stronger the lower the salt concentration is.
%   From a concentration dependent value of $N_\mathrm{FD}$ onwards, the friction per monomer becomes a constant
%   value that increases with increasing salt concentration (indicated by dashed lines).}
%   \label{fig:gammaeff}
% \end{center}
% \end{figure}

In Figure~\ref{fig:gammaeffa}, the effective charge $\Gamma_\mathrm{eff}$ is
displayed as a function of chain length $N$ for different salt concentrations
$c_\mathrm{s}$. The friction increases monotonically with chain length.

Neglecting the contribution of the counterions the effective friction can be
obtained from the hydrodynamic radius $R_\mathrm{h}$ of the polyelectrolyte
defined by:
\[
  \left< \frac{1}{R_\mathrm{h}} \right> = \frac{1}{N} \sum_{i \neq j} \left<
  \frac{1}{\|\vec{r}_i-\vec{r}_j\|} \right>.
\]
Here, $\vec{r}_{i}$ is the position of the $i$-th chain monomer, and
$\vec{r}_{\mathrm{cm}}$ the center of mass of the polyelectrolyte chain. The
angular brackets $\langle \ldots \rangle$ indicate an ensemble average. The
hydrodynamic radius is expected to exhibit a power law scaling $R_\mathrm{h} \sim
\left(N-1\right)^\nu$, where the scaling exponent $\nu$ depends on the system.
For an uncharged polymer with ideal chain behaviour one should get $\nu \approx
0.588$ (Flory exponent)~\cite{moore78a}, whereas for a fully charged
polyelectrolyte without electrostatic screening one expects $\nu = 1$.
Depending on the salt concentration, we obtain values between $\nu \approx
0.66$ for $c_\mathrm{s} = 0 \mathrm{~mM}$ and $\nu \approx 0.59$ for
$c_\mathrm{s} = 160 \mathrm{~mM}$ (not shown here).

Initially, the friction increases with $N$ as given by the hydrodynamic size of the
polyelectrolyte
\[
  \Gamma = 6 \pi \eta R_\mathrm{h} \propto N^{0.61}.
\]
With the onset of counterion condensation the friction exceeds the value of the
bare polyelectrolyte and for long chains
becomes linear in $N$. The higher the concentration of the additional salt,
the earlier the transition between the two regimes is observed. We furthermore note that
the absolute friction value is increased with the addition of external salt.

The role of the additional salt can be best understood when looking at the
effective friction per monomer as presented in Figure~\ref{fig:gammaeffb}.
$\Gamma_\mathrm{eff}/N$ shows an initial decrease with chain length which can be
understood by hydrodynamic shielding: the monomers of short polyelectrolyte
chains are hydrodynamically coupled and shield each other from the effect of the
solvent. This reduces the friction per monomer below the value of a single
particle. The decrease in friction due to the hydrodynamic shielding is stronger
the lower the salt concentration is. The presence of ions in the vicinity of the
chain monomers reduces the hydrodynamic coupling. For longer length scale, \ie
for longer chains, the ions effectively decouple different parts of the
polyelectrolyte chain such that the friction per monomer becomes a constant
value. The chain length $N_\mathrm{FD}$ for which this transition occurs is
depending on the salt concentration. The higher the salt concentration, \ie the
shorter the Debye length in the system, the more confined is the hydrodynamic
shielding effect and the earlier the effective friction becomes constant.

% \begin{figure}[htp]
% \begin{center}
%   \subfloat[]{
%     \includegraphics[width=0.45\columnwidth]{gammaeff-nohd}
%     \label{fig:gammaeff-nohd-a}
%   }
%   \subfloat[]{
%     \includegraphics[width=0.45\columnwidth]{gammaeffN-nohd}
%     \label{fig:gammaeff-nohd-b}
%   }
%   \caption{(a) The normalized effective friction $\Gamma_\mathrm{eff} \mu_1$ as a function of
%   chain length $N$ for different salt concentrations $c_\mathrm{s}$
%   without long-range hydrodynamic interactions. Initially, the increase
%   of the effective friction is super linear, but linear scaling
%   (dotted line) is reached for longer chains. For these chains, the absolute
%   friction value is independent of the addition of external salt.
%   (b) The normalized effective friction per monomer $\Gamma_\mathrm{eff}\mu_1/N$
%   shows an initial increase with chain length that is stronger the
%   higher the salt concentration is. For longer chains, a plateau value
%   is reached which is independent of the salt concentration and can be compared
%   to the predicted value based on the counterion condensation theory
%   (dashed line).}
%   \label{fig:gammaeff-nohd}
% \end{center}
% \end{figure}

The role of hydrodynamic interactions for the effective friction of the
polyelectrolyte can be seen by comparing Figure~\ref{fig:gammaeff} to
Figure~\ref{fig:gammaeff-nohd} that shows the effective friction obtained with
the Langevin algorithm neglecting long-range hydrodynamic interactions. In
Figure~\ref{fig:gammaeff-nohd-a}, the initial increase of the effective friction
is super linear, but linear scaling is reached for longer chains. The absolute
friction value for long chains is independent of the salt concentration. This
can be understood by realizing that the total effective friction of a
polyelectrolyte in the Langevin algorithm is only based on the local friction
parameter $\Gamma_0$ and the number of co-moving particles, \ie the sum of $N$
monomers and $N_\mathrm{CI}$ condensed counterions:
$\Gamma_\mathrm{eff} = \Gamma_0 \left( N + N_\mathrm{CI} \right)$.
As shown in Figure~\ref{fig:qeff} the effective charge, and therefore also
$N_\mathrm{CI}$ and $\Gamma_\mathrm{eff}$, is only influenced by the salt
concentration for short and intermediate chains, but not for long chains.
Figure~\ref{fig:gammaeff-nohd-b} shows the increase of the effective friction
per monomer from $\Gamma_\mathrm{eff}(1) = 1/\mu_1 \approx \Gamma_0$ to a
constant value for long chains, which is comparable to the plateau value
predicted using counterion condensation theory:
$\Gamma_\mathrm{eff}/N = \Gamma_0 \left( 2 - 1/\xi \right)$.

% \begin{figure}[htp]
% \begin{center}
%   \includegraphics[width=\columnwidth]{hi-screening}
%   \caption{Illustration of the influence of surrounding
%   ions on to the long-range hydrodynamic interactions between different parts of
%   a polyelectrolyte chain. a) For an uncharged polymer, the hydrodynamic
%   interactions are unscreened and all chain monomers can interact with each
%   other. b) The presence of counterions during electrophoresis of polyelectrolytes limits
%   the hydrodynamic interaction. c) The more salt is added to the
%   system, the higher is the ion density in the vicinity of the chain,
%   which reduces the hydrodynamic interaction range even further, so
%   that most parts of the chain appear to be hydrodynamically
%   decoupled.}
%   \label{fig:effectivecharge-hiscreening}
% \end{center}
% \end{figure}

Figure~\ref{fig:effectivecharge-hiscreening} schematically illustrates how counterions and
salt in the vicinity of polyelectrolyte chains influence the hydrodynamic interactions during
electrophoresis.  Figure~\ref{fig:effectivecharge-hiscreening}a indicates the regime, where all
parts of the chain can interact via hydrodynamic interactions. The individual
chain segments provide hydrodynamic shielding to each other. During electrophoresis,
see Figure~\ref{fig:effectivecharge-hiscreening}, the
counterions within the polyelectrolyte limit the range of the hydrodynamic interaction.
The hydrodynamic screening length depends on the ion concentration in the vicinity of the chain.
This relation between the ion density and the hydrodynamic screening length was previously
suggested by different authors~\cite{long01a,tanaka01a}.

The connection between electrostatic screening and hydrodynamic screening can
be easily motivated by the following reasoning: the Debye length is the
length-scale on which the charge of the polyelectrolyte is screened by the
surrounding ions. When looking at this object from the outside, the total force
exerted by the applied electric field is zero, \ie no momentum is transferred to
the polyelectrolyte-ion complex. Due to momentum conservation, the interaction
with the fluid has to result in a vanishing total force.

The counterions that associate with the polyelectrolyte influence the solvent
flow around it, effectively canceling the beneficial shielding effects. When
additional salt is added to the system, the like charged salt ions likewise
contribute to this effect as shown in
Figure~\ref{fig:effectivecharge-hiscreening}c. The higher the ion concentration
is, \ie the shorter the Debye length is, the shorter is the length scale on which
different polyelectrolyte monomers can interact hydrodynamically. On a length
scale comparable to the Debye length in the system, different parts of the
polyelectrolyte become decoupled. For longer chains, the effective friction per
segment does not depend on the length of the polyelectrolyte anymore.
Consequently, the effective friction per monomer becomes independent of the
length of the polyelectrolyte chain, as seen in
Figure~\ref{fig:gammaeffb}.

\section{Conclusion}\label{sec:conclusions}

We presented a detailed study of the electrophoretic behaviour of
flexible polyelectrolyte chains by means of a mesoscopic coarse-grained
molecular dynamics model including full hydrodynamic and electrostatic
interactions.

The electrophoretic mobility exhibits a characteristic length dependence for
short polyelectrolyte chains and a constant length independent value for long
chains. We showed that both, the shape and the constant long chain value, depend
on the salt concentration of the solution if hydrodynamical interactions were
properly accounted for. The long chain mobility was then found to be decreasing
with increasing salt concentration, in agreement with experimental observations.

Direct measurements of the effective charge by two independent estimators
showed that the dynamic effective charge for long chains is independent of the
salt concentration. We therefore conclude, that the dependence of the long chain
mobility on the salt concentration is not due a reduced effective charge but has
to be attributed to a change in the effective friction. On the other hand, the
effective charge for short and intermediate chains is influenced by the salt
concentration which explains the different behaviour of the electrophoretic
mobility in this regime.

We note that a static estimate of the effective charge shows a dependence on
the salt concentration leading to to different charge estimates for finite salt
concentrations: a static and a dynamic effective charge.

We showed that the effective friction of the polyelectrolyte is strongly
influenced by the presence of ions in the solution. For short chains and low salt
concentrations no counterions are associated with the polyelectrolyte chain and
the effective friction is given by the hydrodynamic radius. The presence of ions
in the vicinity of the chain reduces the hydrodynamic shielding between the chain
monomers and leads to an increased friction. The longer the chains and the higher
the salt concentration, the more the shielding is reduced, until for chains
longer than a specific length $N_\mathrm{FD}$ the friction becomes linear with
chain length. In this regime, different parts of the chain are effectively
decoupled.

From this, the specific behaviour of the electrophoretic mobility as observed in
experiments can be understood: the hydrodynamic shielding between the monomers
allows for an initial increase in the mobility. The onset of counterion
condensation counteracts this increase as it reduces the effective charge and at
the same time increases the effective friction. For long chains, charge and
friction both become linearly dependent on chain length which therefore results
in the well-known constant mobility, or free-draining limit.

The presence of salt reduces the length scale on which the chain monomers can
interact hydrodynamically. This reduces the initial hydrodynamic shielding and
therefore suppresses the mobility maximum. At the same time, the total friction
is increased leading to a reduced long-chain mobility. Salt concentrations
exceeding the ones in this simulation can cause a total decoupling of the
individual chain monomers, which can then be simulated without hydrodynamic
interactions. We expect a length-dependence of the mobility as shown in
Figure~\ref{fig:mobility-nohd} and an effective friction per monomer,
\cf~Figure~\ref{fig:gammaeffb}, that does not depend on $N$.

The study shows that chemical details and fluid structure can be neglected, and a
higher level of abstraction yields an accurate description of the physics of the
problem, as long as electrostatic and hydrodynamic interactions between all
entities in the system, i.e., the polyelectrolyte, dissociated counterions,
additional salt and the solvent, are properly accounted for. In this way we were able to model
a process bridging the single molecule regime of a few nm up to macromolecules
with contour lengths of more than 100 nm, a regime currently not accessible to
atomistic simulations.

\section*{Acknowledgements}

Funds from the the Volkswagen foundation, the DAAD, and DFG under the TR6 are
gratefully acknowledged. All simulations were carried out on the compute cluster
of the Center for Scientific Computing (CSC) at Goethe University Frankfurt/Main.

%The references should start on their own page.

% BibTeX users can use a .bst file for Faraday Discuss., which can be found at
% http://www.rsc.org/Publishing/ReSourCe/AuthorGuidelines/ElectronicFiles/Templates/tex.asp

%\bibliographystyle{fd-bibtex}
%\bibliography{simbio}

%Please compile a list of all figure captions on a separate page:

\clearpage

\begin{figure}[htp]
\begin{center}
  \includegraphics[width=\columnwidth]{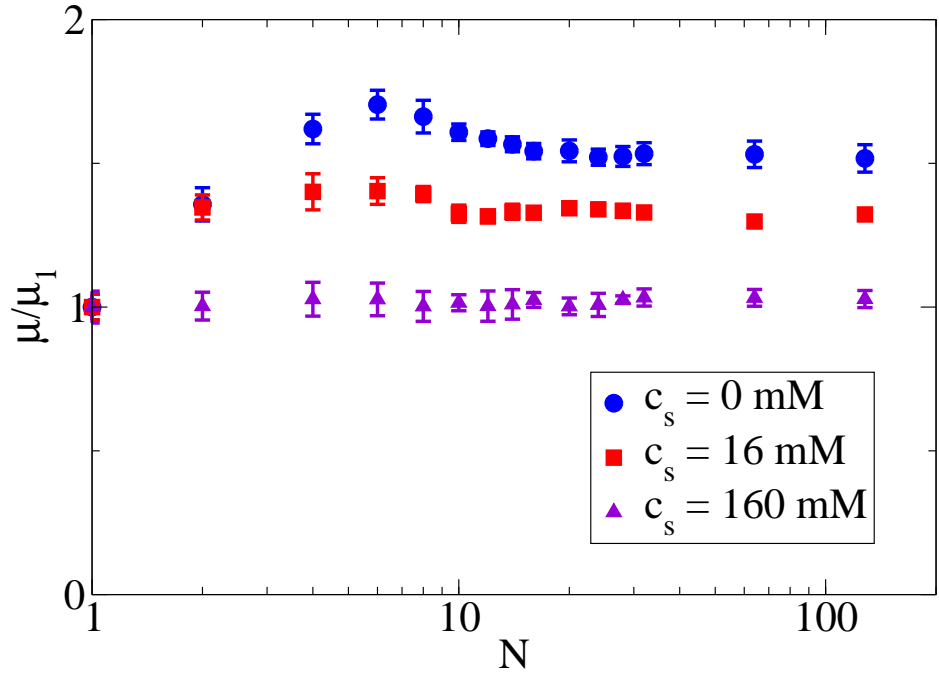}
  \caption{The normalized electrophoretic mobility $\mu/\mu_1$ of polyelectrolyte chains of length
  $N$ for three different salt concentrations using the LB algorithm. The added salt not only influences
  the absolute mobility, but likewise changes the characteristic shape of the
  mobility with respect to chain length $N$.}
  \label{fig:mobility-hd}
\end{center}
\end{figure}

\begin{figure}[htp]
\begin{center}
  \includegraphics[width=\columnwidth]{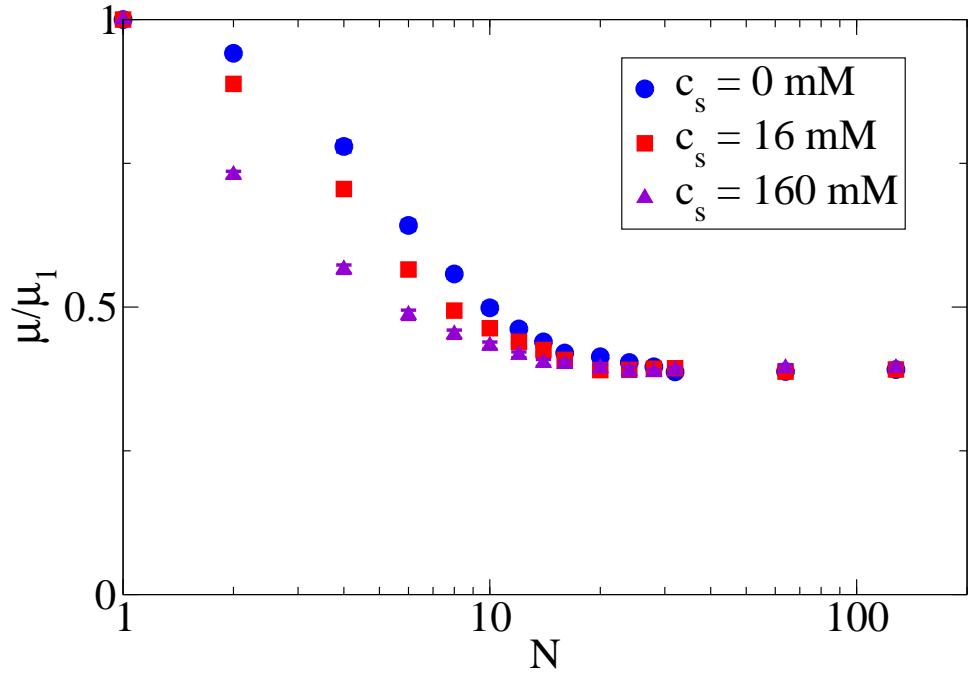}
  \caption{The normalized electrophoretic mobility $\mu/\mu_1$ for different chain
  length $N$ at varying salt concentrations $c_\mathrm{s}$ without hydrodynamic
  interactions differs significantly from the behaviour observed in
  Figure~\ref{fig:mobility-hd}. The mobility shows a salt-dependent monotonic
  decrease for short chains and a salt-independent constant value for long
  chains.}
  \label{fig:mobility-nohd}
\end{center}
\end{figure}

\begin{figure}[htp]
\begin{center}
  \includegraphics[width=\columnwidth]{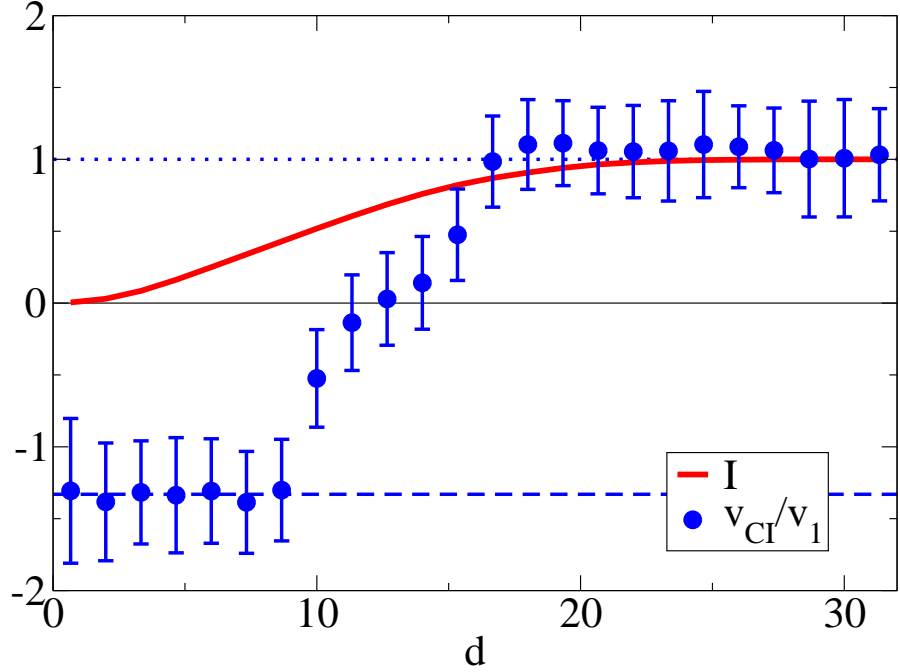}
  \caption{The average ion velocity in the direction of the electric field $v_\mathrm{CI}$ (here for
  a chain with $N=64$ monomers, salt concentration $c_\mathrm{s}=16\mathrm{~mM}$, and hydrodynamics included)
  depends on the distance $d$ to the center of mass of the polyelectrolyte. Ions
  close to the center co-move with the chain's velocity
  (dashed line), whereas ions far away from the center move with the single particle
  velocity $v_1 = \mu_1 E$ into the opposite direction. The distance $d_0$ at which
  $v_\mathrm{CI}\left(d_0\right)=0$ is used to separate co-moving, associated
  ions from non-associated ones.
  The solid line shows the integrated fraction of charges $I$ that is found up to
  the distance $d$ of the center of mass.}
  \label{fig:velcor}
\end{center}
\end{figure}

\begin{figure}[htp]
\begin{center}
  \subfloat[]{
    \includegraphics[width=0.45\columnwidth]{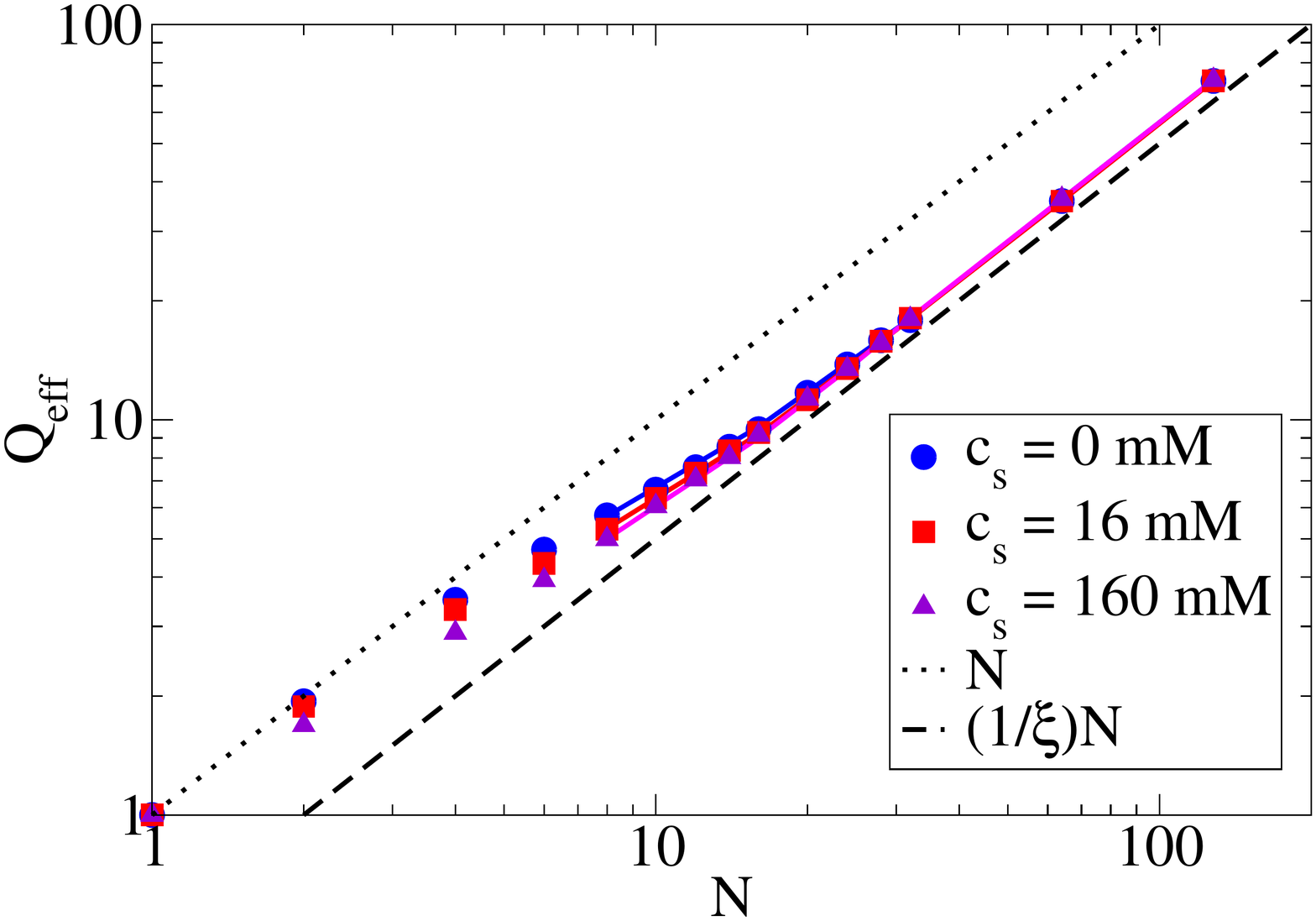}
    \label{fig:qeffa}
  }
  \subfloat[]{
    \includegraphics[width=0.45\columnwidth]{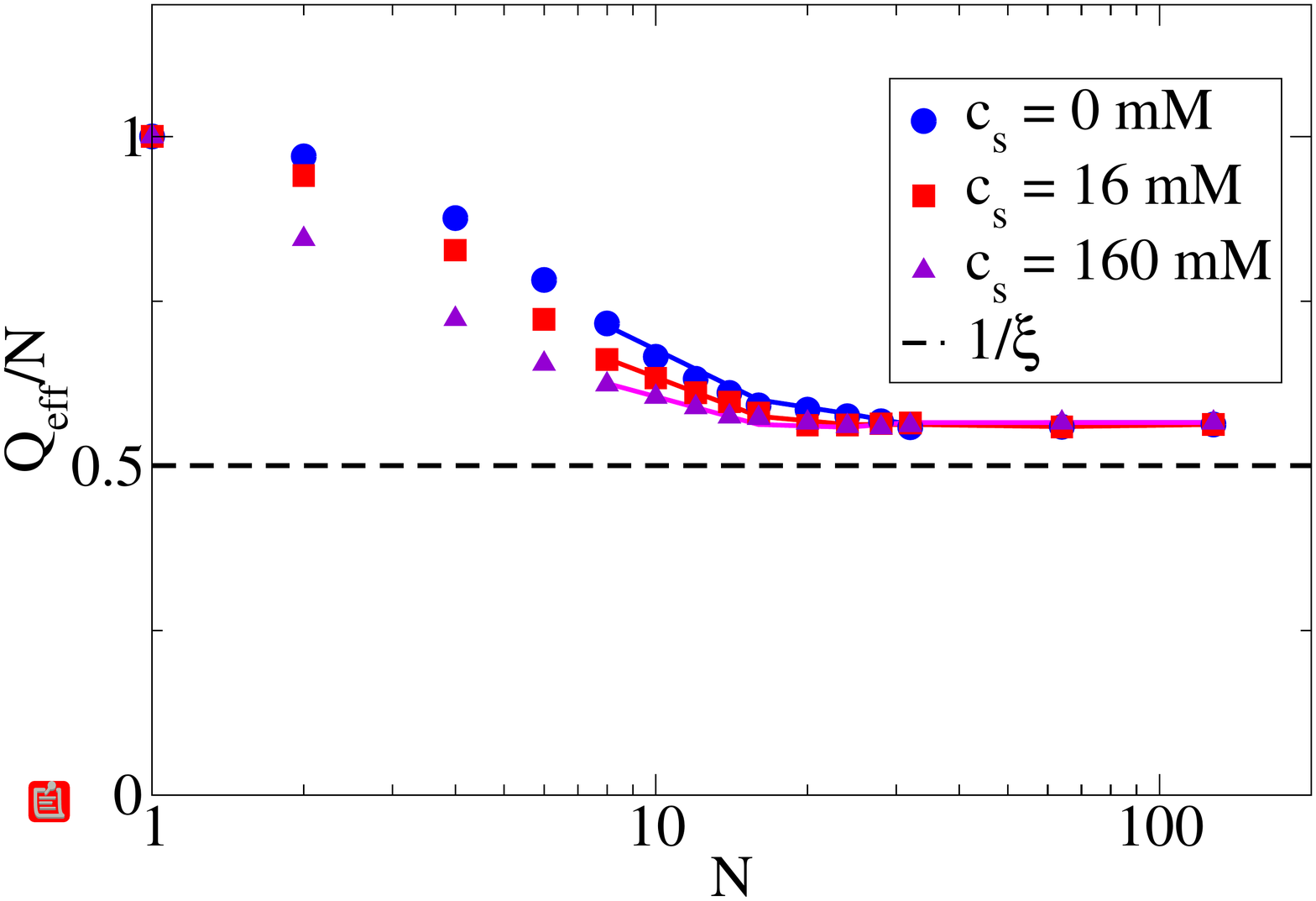}
    \label{fig:qeffb}
  }
  \caption{(a) The effective charge $Q_\mathrm{eff}$ as a function of chain
   length $N$ (symbols for $Q_\mathrm{eff}^{(1)}$, lines for $Q_\mathrm{eff}^{(2)}$).
   Both charge estimators show good agreement. Initially, $Q_\mathrm{eff}$ is
   close to the bare charge $N$ (dotted line), but as ion condensation sets in,
   the effective charge is reduced. Longer
   chains show a linear increase of their charge close to the Manning prediction
   $\left( {1}/{\xi}\right) N$ (dashed line).
   (b) The effective charge per monomer $Q_\mathrm{eff/N}$ is influenced by
   the salt concentration for short and intermediate chains. The higher the
   concentration of the added salt, the faster the electric charge of the
   polyelectrolyte is reduced by condensed ions. For long chains, the charge
   per monomer is again independent of the salt concentration and comparable to
   the Manning prediction $1/\xi$ (dashed line).}
  \label{fig:qeff}
\end{center}
\end{figure}

\begin{figure}[htp]
\begin{center}
  \includegraphics[width=\columnwidth]{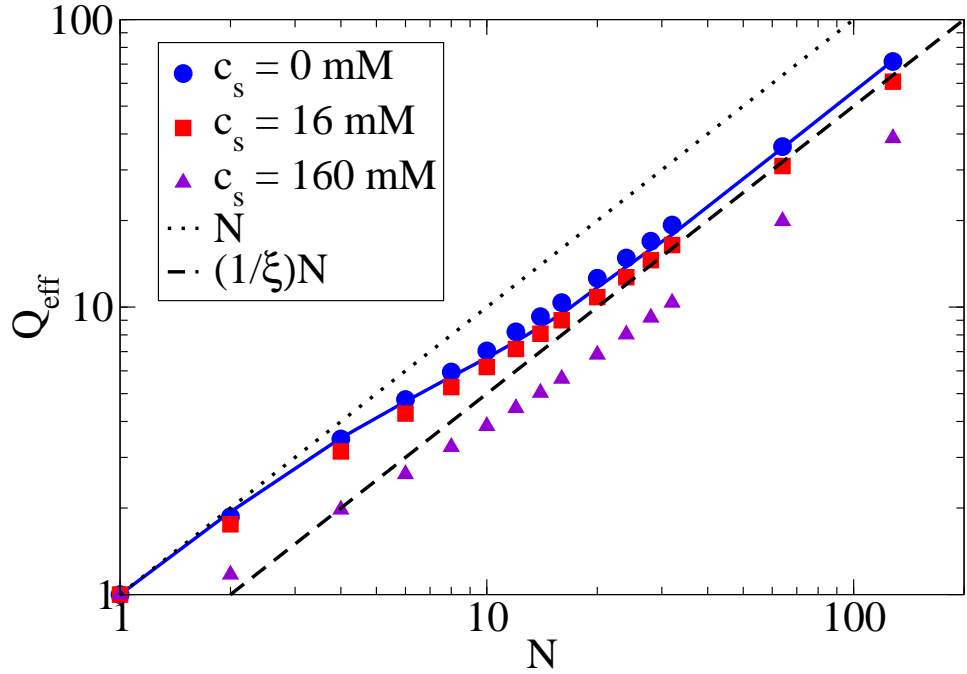}
  \caption{The effective charge as measured by the static estimator
  $Q_\mathrm{eff}^{(3)}$ shows a strong dependence on the the salt
  concentration. At $c_\mathrm{S} = 0 \mathrm{~mM}$ the static estimate
  agrees with the dynamic estimates (solid line). The higher the salt
  concentration, the lower is the static charge estimate. For comparison the
  bare charge $N$ (dotted line) and the Manning prediction (dashed-line) are plotted.}
  \label{fig:qeffstatic}
\end{center}
\end{figure}

\begin{figure}[htp]
\begin{center}
  \subfloat[]{
    \includegraphics[width=0.45\columnwidth]{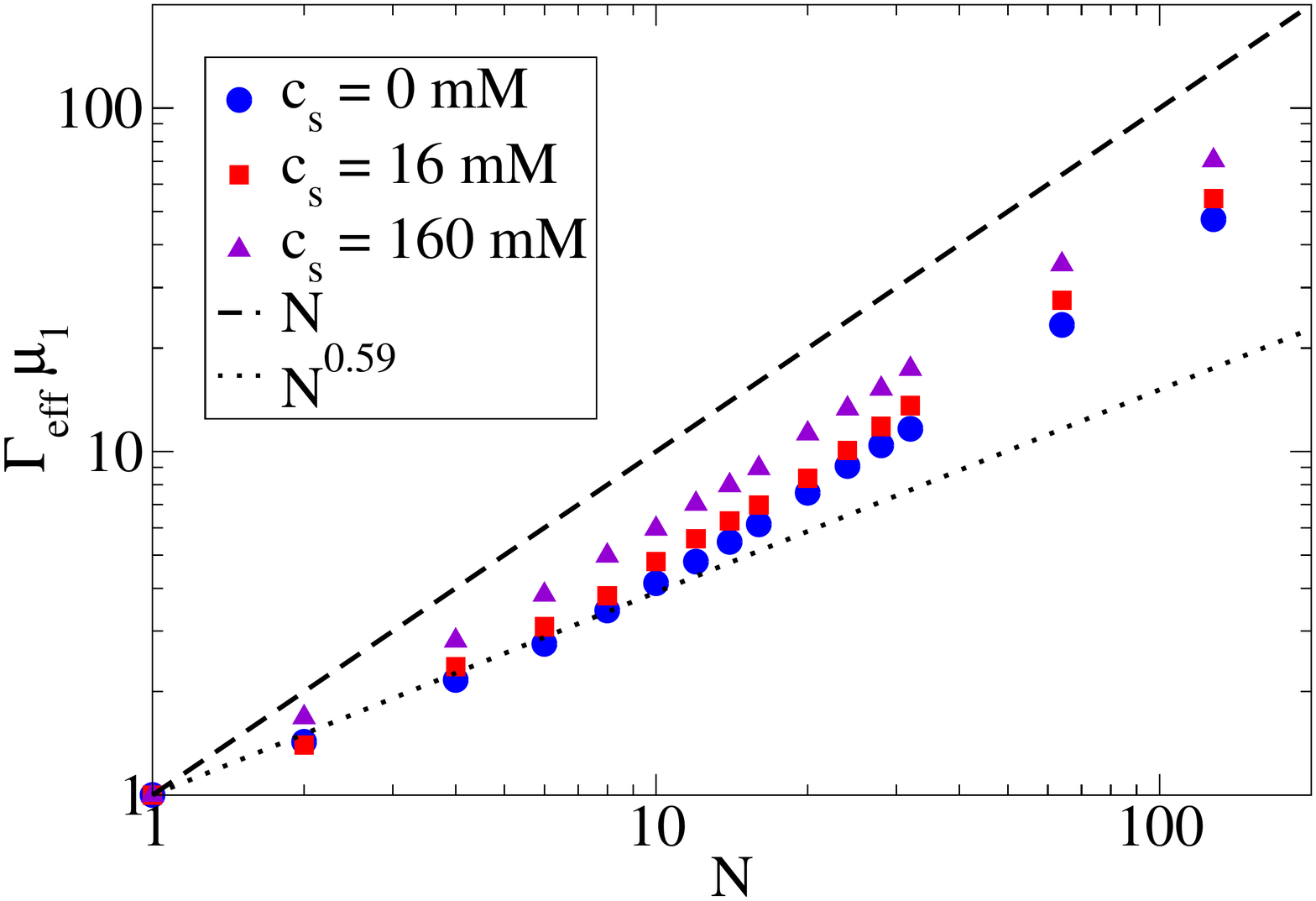}
    \label{fig:gammaeffa}
  }
  \subfloat[]{
    \includegraphics[width=0.45\columnwidth]{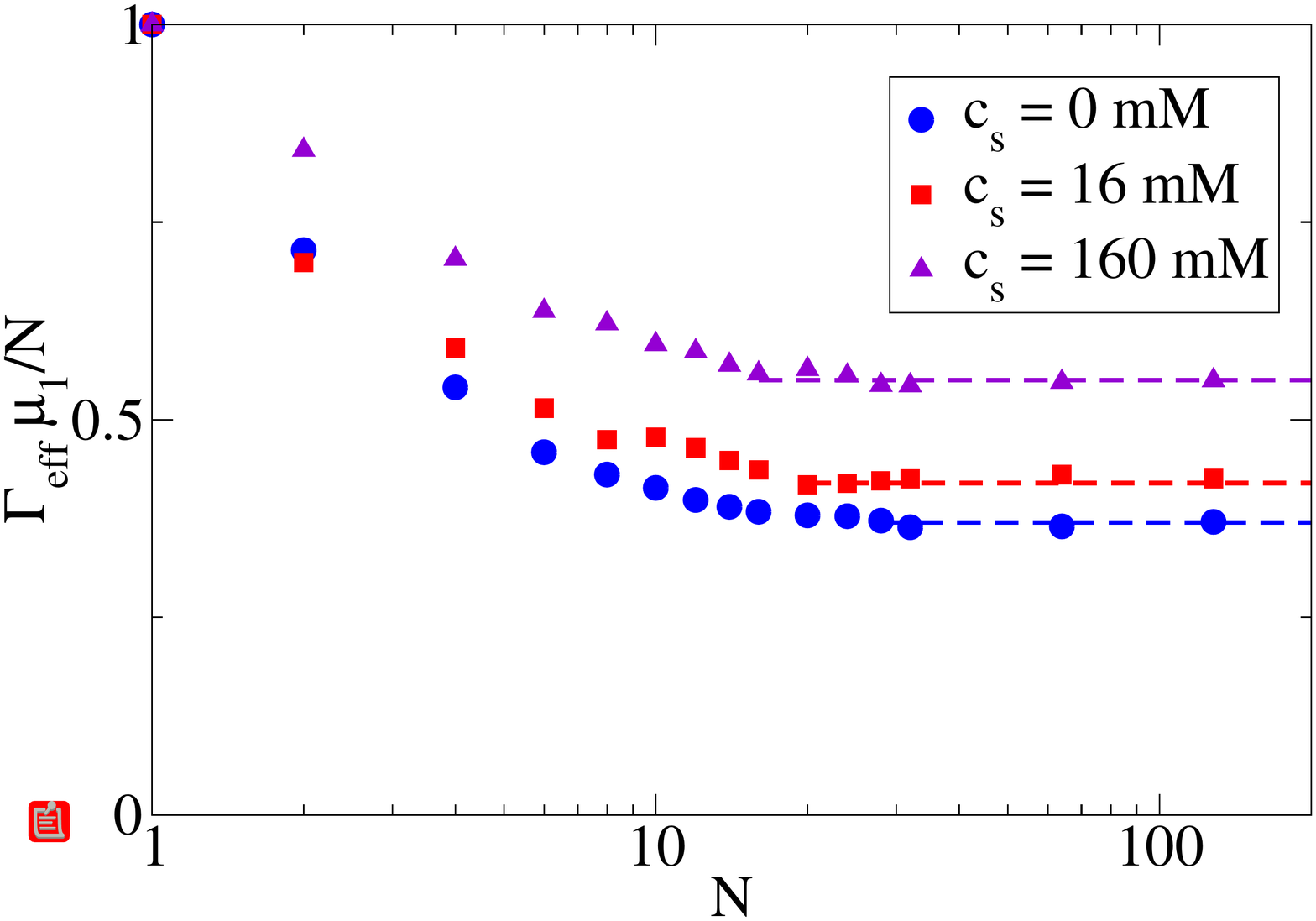}
    \label{fig:gammaeffb}
  }
  \caption{(a) The normalized effective friction $\Gamma_\mathrm{eff} \mu_1$ as a function of
  chain length $N$ for different salt concentrations $c_\mathrm{s}$ using the LB algorithm. Initially,
  the friction increases as given by the hydrodynamic size of the polyelectrolyte
  $\Gamma = \sim N^{0.59}$ (dotted line). With the onset of counterion
  condensation the friction exceeds the value of the bare polyelectrolyte
  and for long chains becomes linear in
  $N$ (dashed line). The absolute friction value is increased with the addition of external salt.
  (b) The normalized effective friction per monomer $\Gamma_\mathrm{eff}\mu_1/N$
  shows an initial decrease with chain length that is stronger the lower the salt concentration is.
  From a concentration dependent value of $N_\mathrm{FD}$ onwards, the friction per monomer becomes a constant
  value that increases with increasing salt concentration (indicated by dashed lines).}
  \label{fig:gammaeff}
\end{center}
\end{figure}

\begin{figure}[htp]
\begin{center}
  \subfloat[]{
    \includegraphics[width=0.45\columnwidth]{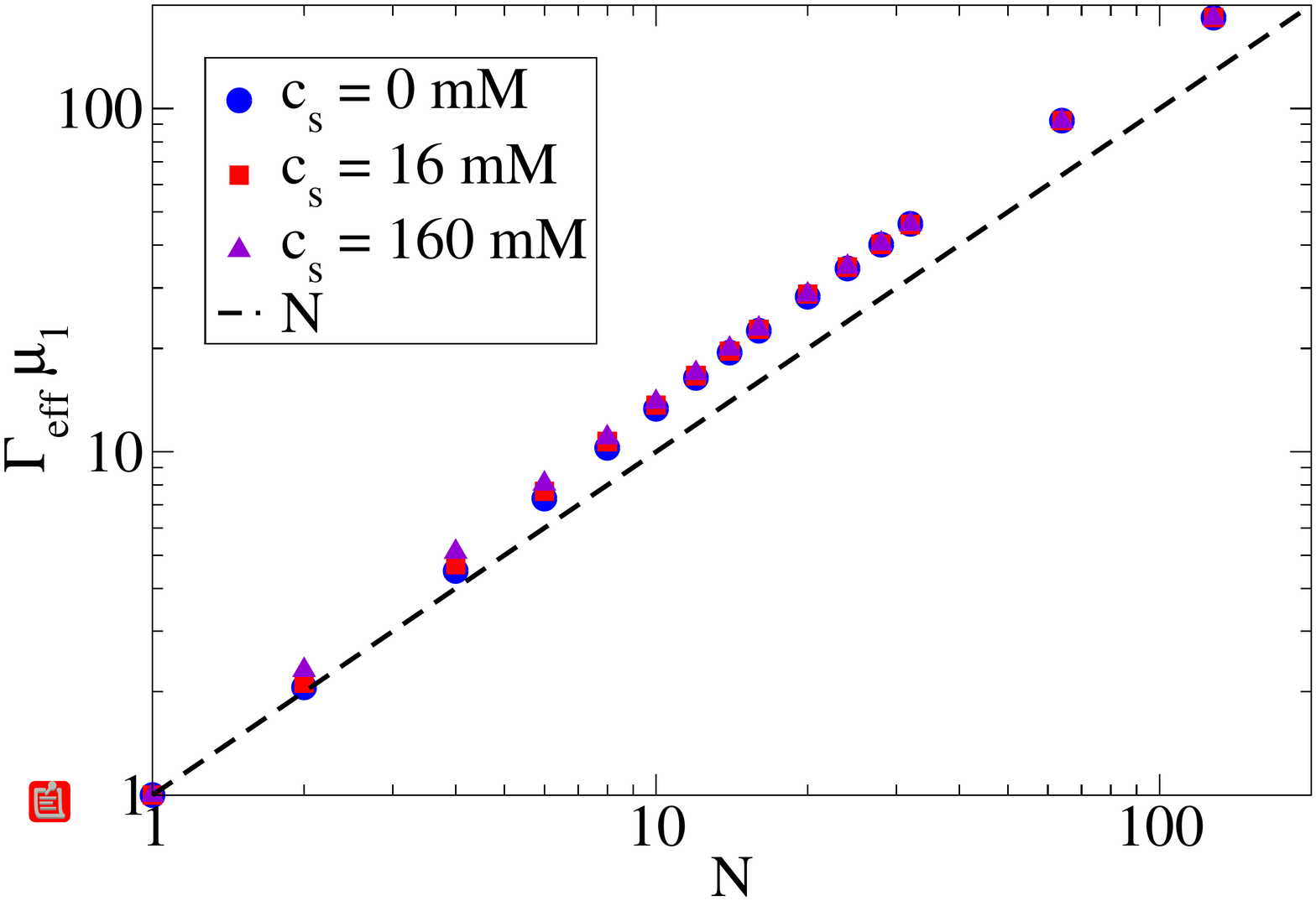}
    \label{fig:gammaeff-nohd-a}
  }
  \subfloat[]{
    \includegraphics[width=0.45\columnwidth]{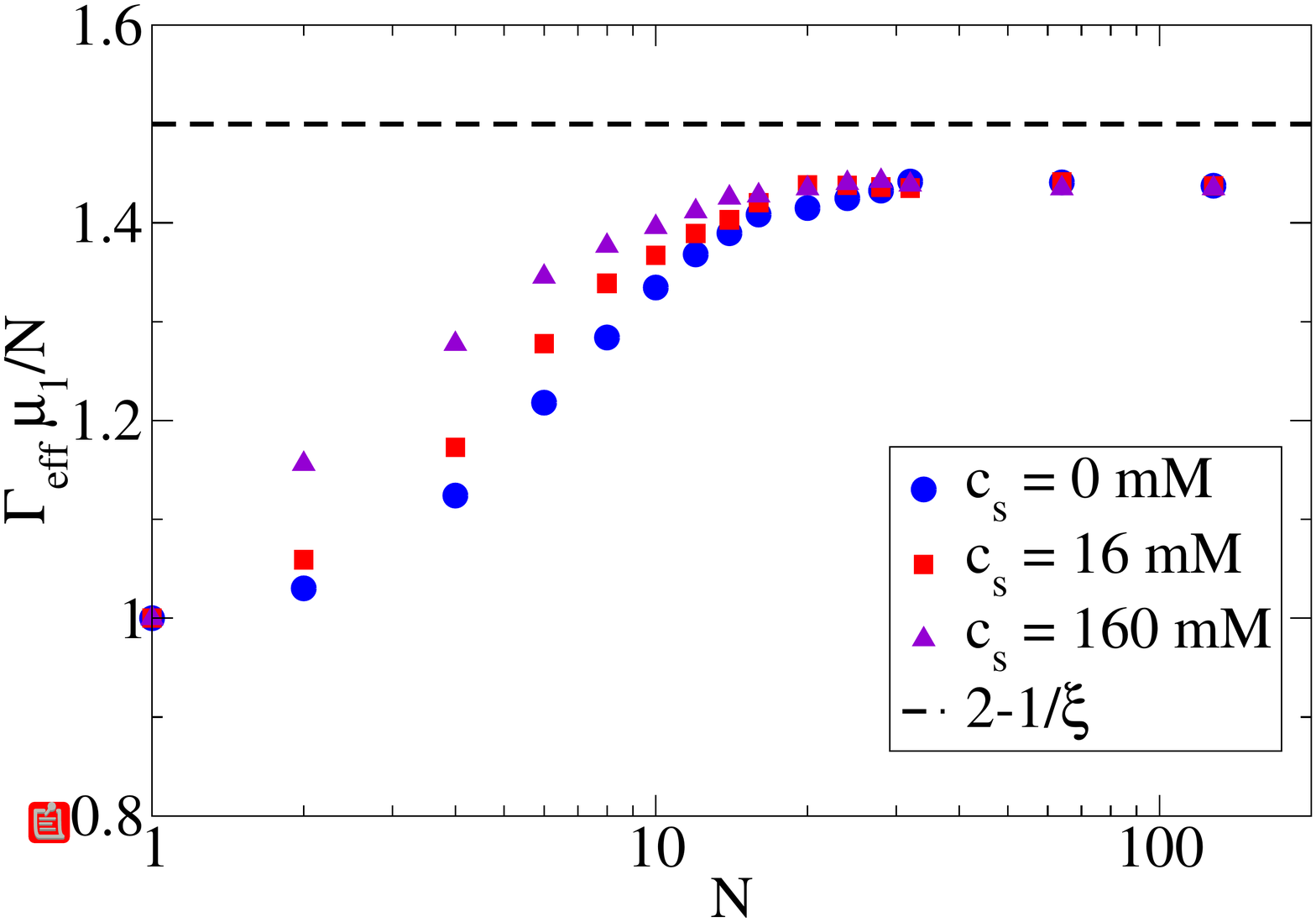}
    \label{fig:gammaeff-nohd-b}
  }
  \caption{(a) The normalized effective friction $\Gamma_\mathrm{eff} \mu_1$ as a function of
  chain length $N$ for different salt concentrations $c_\mathrm{s}$
  without long-range hydrodynamic interactions. Initially, the increase
  of the effective friction is super linear, but linear scaling
  (dotted line) is reached for longer chains. For these chains, the absolute
  friction value is independent of the addition of external salt.
  (b) The normalized effective friction per monomer $\Gamma_\mathrm{eff}\mu_1/N$
  shows an initial increase with chain length that is stronger the
  higher the salt concentration is. For longer chains, a plateau value
  is reached which is independent of the salt concentration and can be compared
  to the predicted value based on the counterion condensation theory
  (dashed line).}
  \label{fig:gammaeff-nohd}
\end{center}
\end{figure}

\begin{figure}[htp]
\begin{center}
  \includegraphics[width=\columnwidth]{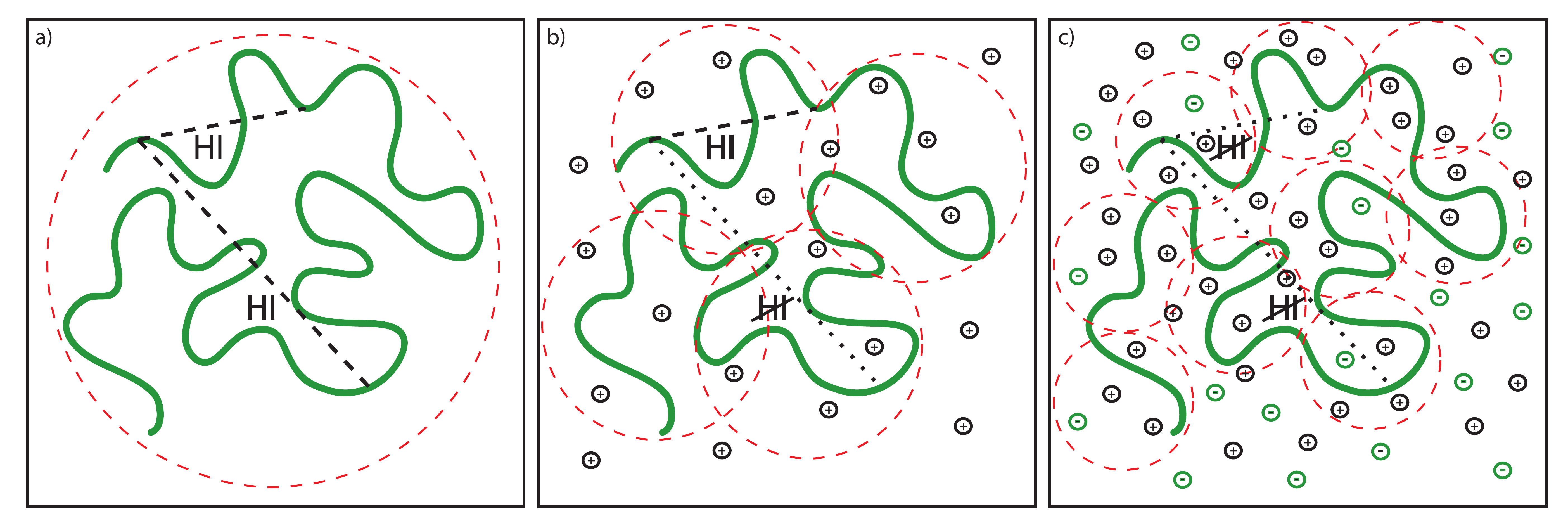}
  \caption{Illustration of the influence of surrounding
  ions on to the long-range hydrodynamic interactions between different parts of
  a polyelectrolyte chain. (a) For an uncharged polymer, the hydrodynamic
  interactions are unscreened and all chain monomers can interact with each
  other. (b) The presence of counterions during electrophoresis of
  polyelectrolytes limits the hydrodynamic interaction. (c) The more salt is
  added to the system, the higher is the ion density in the vicinity of the chain,
  which reduces the hydrodynamic interaction range even further, so
  that most parts of the chain appear to be hydrodynamically
  decoupled.}
  \label{fig:effectivecharge-hiscreening}
\end{center}
\end{figure}

\begin{list}{}{\leftmargin 2cm \labelwidth 1.5cm \labelsep 0.5cm}

\item[\bf Fig. 1] The normalized electrophoretic mobility
    $\mu / \mu_{\mathrm{FD}}$ as a function of the number of repeat
    units $N$ for simulation data including hydrodynamic interactions (HI),
    and experimental data coming from capillary electrophoresis (CE) and from
    electrophoretic NMR.  The inset compares to simulation data obtained with a
    model neglecting hydrodynamic interactions.

\item[\bf Fig. 2] The normalized electrophoretic mobility $\mu/\mu_1$ of polyelectrolyte chains of length
  $N$ for three different salt concentrations using the LB algorithm. The added salt not only influences
  the absolute mobility, but likewise changes the characteristic shape of the
  mobility with respect to chain length $N$.
  
\item[\bf Fig. 3] The normalized electrophoretic mobility $\mu/\mu_1$ for different chain
  length $N$ at varying salt concentrations $c_\mathrm{s}$ without hydrodynamic
  interactions differs significantly from the behaviour observed in
  Figure~\ref{fig:mobility-hd}. The mobility shows a salt-dependent monotonic
  decrease for short chains and a salt-independent constant value for long
  chains.

\item[\bf Fig. 4] The average ion velocity in the direction of the electric
field $v_\mathrm{CI}$ (here for a chain with $N=64$ monomers, salt concentration $c_\mathrm{s}=16\mathrm{~mM}$, and hydrodynamics included)
  depends on the distance $d$ to the center of mass of the polyelectrolyte. Ions
  close to the center co-move with the chain's velocity
  (dashed line), whereas ions far away from the center move with the single particle
  velocity $v_1 = \mu_1 E$ into the opposite direction. The distance $d_0$ at which
  $v_\mathrm{CI}\left(d_0\right)=0$ is used to separate co-moving, associated
  ions from non-associated ones.
  The solid line shows the integrated fraction of charges $I$ that is found up to
  the distance $d$ of the center of mass.
   
\item[\bf Fig. 5] (a) The effective charge $Q_\mathrm{eff}$ as a function of
chain length $N$ (symbols for $Q_\mathrm{eff}^{(1)}$, lines for $Q_\mathrm{eff}^{(2)}$).
   Both charge estimators show good agreement. Initially, $Q_\mathrm{eff}$ is
   close to the bare charge $N$ (dotted line), but as ion condensation sets in,
   the effective charge is reduced. Longer
   chains show a linear increase of their charge close to the Manning prediction
   $\left( {1}/{\xi}\right) N$ (dashed line).
   (b) The effective charge per monomer $Q_\mathrm{eff/N}$ is influenced by
   the salt concentration for short and intermediate chains. The higher the
   concentration of the added salt, the faster the electric charge of the
   polyelectrolyte is reduced by condensed ions. For long chains, the charge
   per monomer is again independent of the salt concentration and comparable to
   the Manning prediction $1/\xi$ (dashed line).
  
\item[\bf Fig. 6] The effective charge as measured by the static estimator
  $Q_\mathrm{eff}^{(3)}$ shows a strong dependence on the the salt
  concentration. At $c_\mathrm{S} = 0 \mathrm{~mM}$ the static estimate
  agrees with the dynamic estimates (solid line). The higher the salt
  concentration, the lower is the static charge estimate. For comparison the
  bare charge $N$ (dotted line) and the Manning prediction (dashed-line) are plotted.
\clearpage  
\item[\bf Fig. 7] (a) The normalized effective friction $\Gamma_\mathrm{eff}
\mu_1$ as a function of chain length $N$ for different salt concentrations $c_\mathrm{s}$ using the LB algorithm. Initially,
  the friction increases as given by the hydrodynamic size of the polyelectrolyte
  $\Gamma = \sim N^{0.59}$ (dotted line). With the onset of counterion
  condensation the friction exceeds the value of the bare polyelectrolyte
  and for long chains becomes linear in
  $N$ (dashed line). The absolute friction value is increased with the addition of external salt.
  (b) The normalized effective friction per monomer $\Gamma_\mathrm{eff}\mu_1/N$
  shows an initial decrease with chain length that is stronger the lower the salt concentration is.
  From a concentration dependent value of $N_\mathrm{FD}$ onwards, the friction per monomer becomes a constant
  value that increases with increasing salt concentration (indicated by dashed lines).
   
\item[\bf Fig. 8] (a) The normalized effective friction $\Gamma_\mathrm{eff} \mu_1$ as a function of
  chain length $N$ for different salt concentrations $c_\mathrm{s}$
  without long-range hydrodynamic interactions. Initially, the increase
  of the effective friction is super linear, but linear scaling
  (dotted line) is reached for longer chains. For these chains, the absolute
  friction value is independent of the addition of external salt.
  (b) The normalized effective friction per monomer $\Gamma_\mathrm{eff}\mu_1/N$
  shows an initial increase with chain length that is stronger the
  higher the salt concentration is. For longer chains, a plateau value
  is reached which is independent of the salt concentration and can be compared
  to the predicted value based on the counterion condensation theory
  (dashed line).
 
\item[\bf Fig. 9] Illustration of the influence of surrounding
  ions on to the long-range hydrodynamic interactions between different parts of
  a polyelectrolyte chain. (a) For an uncharged polymer, the hydrodynamic
  interactions are unscreened and all chain monomers can interact with each
  other. (b) The presence of counterions during electrophoresis of
  polyelectrolytes limits the hydrodynamic interaction. (c) The more salt is
  added to the system, the higher is the ion density in the vicinity of the chain,
  which reduces the hydrodynamic interaction range even further, so
  that most parts of the chain appear to be hydrodynamically
  decoupled.
 
\end{list}

% Figures
\clearpage

\clearpage

\clearpage

\clearpage

\clearpage

\clearpage

\clearpage

\clearpage

\clearpage

\end{document}